\documentclass[conference]{IEEEtran}
\IEEEoverridecommandlockouts
\usepackage{cite}
\usepackage{amsmath,amssymb,amsfonts}
\usepackage{graphicx}
\usepackage{textcomp}
\usepackage{balance}
\makeatletter
\newcommand{\newlineauthors}{%
  \end{@IEEEauthorhalign}\hfill\mbox{}\par
  \mbox{}\hfill\begin{@IEEEauthorhalign}
}
\makeatother
\def\BibTeX{{\rm B\kern-.05em{\sc i\kern-.025em b}\kern-.08em
    T\kern-.1667em\lower.7ex\hbox{E}\kern-.125emX}}
\begin{document}

\title{Factors Impacting Faculty Adoption of Project-Based Learning in Computing Education: a Survey}

\author{
\IEEEauthorblockN{Ahmad D. Suleiman}
\IEEEauthorblockA{
\textit{Computing and Information Sciences}\\
\textit{Rochester Institute of Technology}\\
Rochester, NY, USA\\
as4300@rit.edu\\
}
\and
\IEEEauthorblockN{Yiming Tang}
\IEEEauthorblockA{
\textit{Department of Software Engineering}\\
\textit{Rochester Institute of Technology}\\
Rochester, NY, USA\\
yxtvse@rit.edu\\
}
\and
\IEEEauthorblockN{Daqing Hou}
\IEEEauthorblockA{
\textit{Department of Software Engineering}\\
\textit{Rochester Institute of Technology}\\
Rochester, NY, USA\\
dqvse@rit.edu\\
}

}

\maketitle

\begin{abstract}
This research full paper investigates the factors influencing computing educators’ adoption of project-based learning (PjBL) in software engineering and computing curricula. Recognized as a student-centered pedagogical approach, PjBL has the potential to enhance student motivation, engagement, critical thinking, collaboration, and problem-solving skills. Despite these benefits, faculty adoption remains inconsistent due to challenges such as insufficient institutional support, time constraints, limited training opportunities,  designing or sourcing projects, and aligning them with course objectives. 
This research explores these barriers and investigates the strategies and resources that facilitate a successful adoption.
Using a mixed-methods approach, data from 80 computing faculty were collected through an online survey comprising closed-ended questions to quantify barriers, enablers, and resource needs, along with an open-ended question to gather qualitative insights. Quantitative data were analyzed using statistical methods, while qualitative responses underwent thematic analysis. 
Results reveal that while PjBL is widely valued, its adoption is often selective and impacted by challenges in planning and managing the learning process, designing suitable projects, and a lack of institutional support, such as time, funding, and teaching assistants. Faculty are more likely to adopt or sustain PjBL when they have access to peer collaboration, professional development, and institutional incentives. In addition, sourcing projects from research, industry partnerships, and borrowing from peers emerged as key facilitators for new projects. These findings underscore the need for systemic support structures to empower faculty to experiment with and scale PjBL practices.

\end{abstract}


\IEEEpeerreviewmaketitle

\section{Introduction}
\label{section:introduction}
PjBL has gained recognition as a student-centered pedagogical approach that enhances student motivation, engagement, critical thinking, collaboration, and problem-solving skills \cite{bell2010project, kokotsaki2016project}. Despite its proven benefits in computing education, faculty adoption remains inconsistent due to challenges such as insufficient institutional incentives and support, difficulties in designing or sourcing projects \cite{gupta2022impact}, aligning projects with course objectives, and implementing the learning process related to project organization \cite{fu2018teaching,fioravanti2018integrating}, providing student support \cite{tubino2021reforming, papadopoulos2012students, suleiman2024providing}, and assessment \cite{daun2016project,fernandes2020achieving}.

Even faculty who have adopted the PjBL approach often face ongoing challenges in maintaining and renewing course projects. Changing or redesigning projects instead of reusing previous ones has several merits. In computing courses, one practical benefit is that it prevents academic dishonesty \cite{mason2019collaboration}. If the same project is reused every semester, students might find solutions from prior offerings or online repositories, undermining the learning process. By introducing new or modified projects, instructors ensure that each cohort of students engages in original work. In addition, keeping projects up-to-date and relevant is a constant concern. Computing technology in real-world contexts evolves quickly. A project that was cutting-edge two years ago might feel stale or mismatched to current industry practices. Therefore, instructors need to refresh projects to maintain relevance and student interest. 

However, constantly inventing new projects comes with its own challenges. Designing a complex project from scratch requires creativity, content expertise, and foresight about possible student difficulties \cite{marti2006pbl}. Faculty must consider scope (ensuring the project is neither too broad nor too narrow), how to support student learning throughout the project, and how to assess student work. Some faculty utilize PjBL projects developed by others. This approach can preserve the effort invested in designing an entirely new project, but it still requires effort to review and scope the project to one's needs.

This study is motivated by the need to understand these challenges and identify factors that drive or hinder them in computing education. The study aims to answer the following research questions:  
\begin{itemize}
    \item \textbf{RQ1:} What barriers prevent computing faculty from adopting the PjBL teaching approach? 
    \item \textbf{RQ2:} What challenges prevent faculty from designing or adopting a new course project to replace existing ones?
    \item \textbf{RQ3:} For faculty who have successfully designed or adopted a new course project, what strategies or resources do they perceive as essential to their success?
\end{itemize}
By addressing these factors, this research contributes to the growing body of literature aimed at improving PjBL teaching practices and educational outcomes in the computing domain.

This research adopts a mixed-methods approach to understand faculty-specific factors influencing PjBL adoption. Data were collected through an online survey targeting computing faculty across higher education institutions. The survey includes (a) closed-ended questions to quantify barriers, enablers, and perceived resource needs. (b) An open-ended question to gather qualitative insights into challenges and strategies for designing or adopting new projects. Quantitative data were analyzed using descriptive and inferential statistical methods, while qualitative data underwent thematic analysis to identify patterns and actionable insights. We received 80 responses from diverse instructors in terms of teaching experience and frequency of PjBL adoption.

The rest of the paper is organized as follows: Section \ref{section:theoritical_framework} outlines the theoretical framework of this study. Section \ref{section:methodology} details the research methodology that guides this research. We present the survey results in Sections \ref{section:pjbl_adoption}, \ref{section:project_design}, and \ref{section:project_adoption}. We discuss the results and highlight the study's limitations in Section \ref{section:discussion}. Finally, we conclude in Section \ref{section:conlcusion}.

\section{Theoretical Framework}
\label{section:theoritical_framework}

This study is grounded in Rogers’ Diffusion of Innovations theory (DOI) \cite{rogers2014diffusion} and draws upon prior research \cite{meng2023tackle, rehman_trends} on PjBL implementation to frame the investigation of faculty adoption factors. Rogers’ DOI theory is a fitting lens here, as it has been widely used to examine educational innovation uptake \cite{sahin2006detailed}. DOI emphasizes that the type of innovation-decision, innovation’s perceived attributes, and social context all shape innovation adoption. 

The type of innovation-decision (e.g., authority-based, collective, or optional) can be directly influenced by institutional directives. Strong institutional mandates may prompt more top-down (authority-based) adoption, while robust support and encouragement can facilitate collective or optional adoptions by reducing perceived risk and empowering faculty choice. 

DOI outlines several innovation attributes that influence an innovation’s adoption: relative advantage, compatibility, complexity (or simplicity), trialability, and observability \cite{rogers2014diffusion}. In the context of PjBL, these factors illuminate faculty adoption behaviors. According to Rogers, innovation is more readily adopted when it addresses users’ needs and challenges \cite{rees2019opportunities}. Consistent with this idea, perceived usefulness and support play a critical role. Faculty are more likely to adopt PjBL if they believe it will solve instructional problems and if they have the necessary support to implement it \cite{rees2019opportunities, meng2023tackle}. Compatibility issues arise when PjBL is misaligned with existing curricula or assessment demands. For example, a misfit between PjBL’s open-ended projects and rigid computing curricula can deter adoption \cite{meng2023tackle}. The complexity of designing and managing projects, especially in fast-evolving computing fields, can make PjBL seem difficult to implement. Instructors may be unsure how to start small (low trialability) or may not readily see peers’ successes (low observability), further slowing down diffusion. 

A key element of DOI is the social system, highlighting that institutional context and support can significantly facilitate (or hinder) innovation adoption \cite{rogers2014diffusion}. In the case of PjBL, institutional culture and support have emerged as pivotal factors influencing faculty adoption decisions. Studies have identified that a lack of support, whether administrative or financial, is a major barrier to PjBL implementation \cite{rehman_trends, meng2023tackle}. Conversely, strong institutional backing can empower faculty to experiment with and sustain PjBL. Kokotsaki et al. \cite{kokotsaki2016project} noted that providing administrative support (e.g., encouragement from leadership, resource allocation) is one of the critical enablers for successful PjBL integration \cite{meng2023tackle}. When faculty receive training, time release, and resources to redesign their courses, they are more willing to embrace PjBL \cite{rehman_trends}. Such support aligns with Rogers’ notion that a supportive social system and organizational climate greatly enhance diffusion.

\section{Research Methodology}
\label{section:methodology}
This section details our research methodology, including questionnaire design, participant recruitment, data analysis, and the background information of the faculty from the survey responses.

\subsection{Research and Questionnaire Design}
A mixed-methods design was employed to capture quantitative and qualitative data, providing a comprehensive view of the factors impacting PjBL adoption. The quantitative component allowed us to measure the prevalence of various barriers and enablers among computing faculty, while the qualitative component provided deeper insights into personal experiences and contextual challenges. Google Forms was chosen as the survey platform because of its simplicity, accessibility, and because it facilitates efficient data collection.

\begin{table*}
\caption{Shortened Version of the Survey Questionnaire}
\begin{center}
\begin{tabular}{|p{5.9in}|c|}
\hline

\textbf{Demographics and Background Questions}& \textbf{Question Type}\\
\hline
How would you rate your personal belief in high-quality teaching?&High/Medium/Low\\
\hline
How would you describe your institution’s emphasis on high-quality teaching of your PjBL course(s)?&High/Medium/Low\\
\hline
What levels of courses do you primarily teach?&Multiple\\
\hline

\textbf{Adoption of PjBL as a Teaching Approach Questions}&\\
\hline
How often have you adopted PjBL in your courses?&Single+Open\\
\hline
In your opinion, what are the primary barriers preventing the adoption of the PjBL approach?&Multiple+Open\\
\hline
In your opinion, what types of support and strategies have/would have encouraged you to adopt the PjBL approach in your classes?&Multiple+Open\\
\hline

\textbf{Designing New PjBL Projects Questions}&\\
\hline
How often do you design new PjBL projects to replace the previous course projects? &Single+Open\\
\hline
In your opinion, what are the primary barriers to designing new projects?&Multiple+Open\\
\hline
In your opinion, what strategies or resources have/would have helped you successfully design PjBL projects?&Multiple+Open\\
\hline

\textbf{Adopting Existing PjBL Projects Questions}&\\
\hline
How often do you adopt new PjBL projects for your courses to replace the previous projects?&Single+Open\\
\hline
In your opinion, what are the primary barriers preventing you from adopting new projects to replace previous ones? &Multiple+Open\\
\hline
In your opinion, what strategies or resources have/would have helped you in adopting new projects to replace previous ones?  &Multiple+Open\\
\hline

\textbf{Open Feedback Question}&\\
\hline
What additional insights or recommendations would you like to share about incorporating PjBL in a classroom,  as well as designing new PjBL projects or adopting existing ones in computing education?  &Open\\
\hline
\end{tabular}
\label{tab:questionnaire}
\end{center}
\end{table*}

The questionnaire was developed through an integrated consideration of our teaching experiences, Rogers's DOI theory \cite{rogers2014diffusion}, and a review of relevant literature on PjBL in computing education. The survey is organized into five sections that address key dimensions of PjBL adoption: demographics and background, adoption of PjBL as a teaching approach, designing new PjBL projects, adopting existing PjBL projects, and open feedback. Table \ref{tab:questionnaire} depicts a concise overview of the questionnaire.

\subsection{Participant Recruitment}

The questionnaire was targeted at computing instructors.  We used three methods of participant recruitment:
\begin{itemize}
    \item Direct email invitation to instructors. The email list comprises 313 recipients obtained from research papers in our ongoing work on a systematic review of computing PjBL \cite{ahmadSLR}. The review gathered 184 computing-related PjBL papers. 
    \item Direct contact with CS/SE departments and referral-based recruitment further broadened the participant pool.
    \item Utilizing online social media platforms like Reddit.
\end{itemize}

This study adheres to strict ethical guidelines. All participation was voluntary and anonymous. Furthermore, the research underwent Institutional Review Board (IRB) approval before data collection, ensuring that all ethical standards for research with human subjects were met. Informed consent was obtained from all participants, outlining their rights and the measures taken to secure confidentiality. All data were reported in aggregate form to protect individual privacy. We received a total of 80 responses. 

\subsection{Data Analysis}
Quantitative data from closed-ended survey questions were analyzed using descriptive statistics to summarize response distributions. To investigate relationships between background information (e.g., demographics, teaching experience, course levels) and key outcomes (adoption, barriers, and facilitators), we performed a chi-square independence test (\(\alpha = 0.05\)) \cite{mchugh2013chi}. The chi-square test allowed us to assess whether categorical background variables significantly affect PjBL adoption, barriers, and facilitators. Given that the qualitative data were derived from open-ended survey responses, we employed a thematic analysis approach following the widely recognized six-phase process outlined by Braun and Clarke \cite{Braun01012006}. Through this procedure, we identified recurring themes that enrich and contextualize the quantitative findings. 

\subsection{Demographics and Background of Responses}
Below are the demographics and backgrounds of the survey respondents:

\subsubsection{Teaching Experience}
The majority of respondents (72.5\%) identify as experienced instructors with over 11 years of teaching experience. A smaller portion, 18.8\%, consider themselves at an intermediate level with 6 to 10 years of experience, while only 8.8\% classify themselves as beginners with 5 or fewer years of teaching experience. This distribution indicates that the survey primarily gathered perspectives from instructors with extensive teaching experience, which may offer valuable insights into the barriers and facilitators influencing the adoption of PjBL in computing education.

\subsubsection{Personal Belief and Institutional Emphasis on High-Quality Teaching}
Nearly all respondents (96.3\%) indicate a strong personal commitment to high-quality teaching, with only three neutral. Meanwhile, approximately two-thirds (66.3\%) feel their institution also places high importance on quality teaching, but 30\% report a neutral stance, and three perceive a low level of institutional emphasis. This contrast suggests that while faculty generally hold strong individual beliefs about teaching excellence, they do not universally perceive their institutions to share the same level of commitment.

\subsubsection{Teaching Grade Level}
\begin{figure}[tb]
    \centering
    \includegraphics[width=0.5\linewidth]{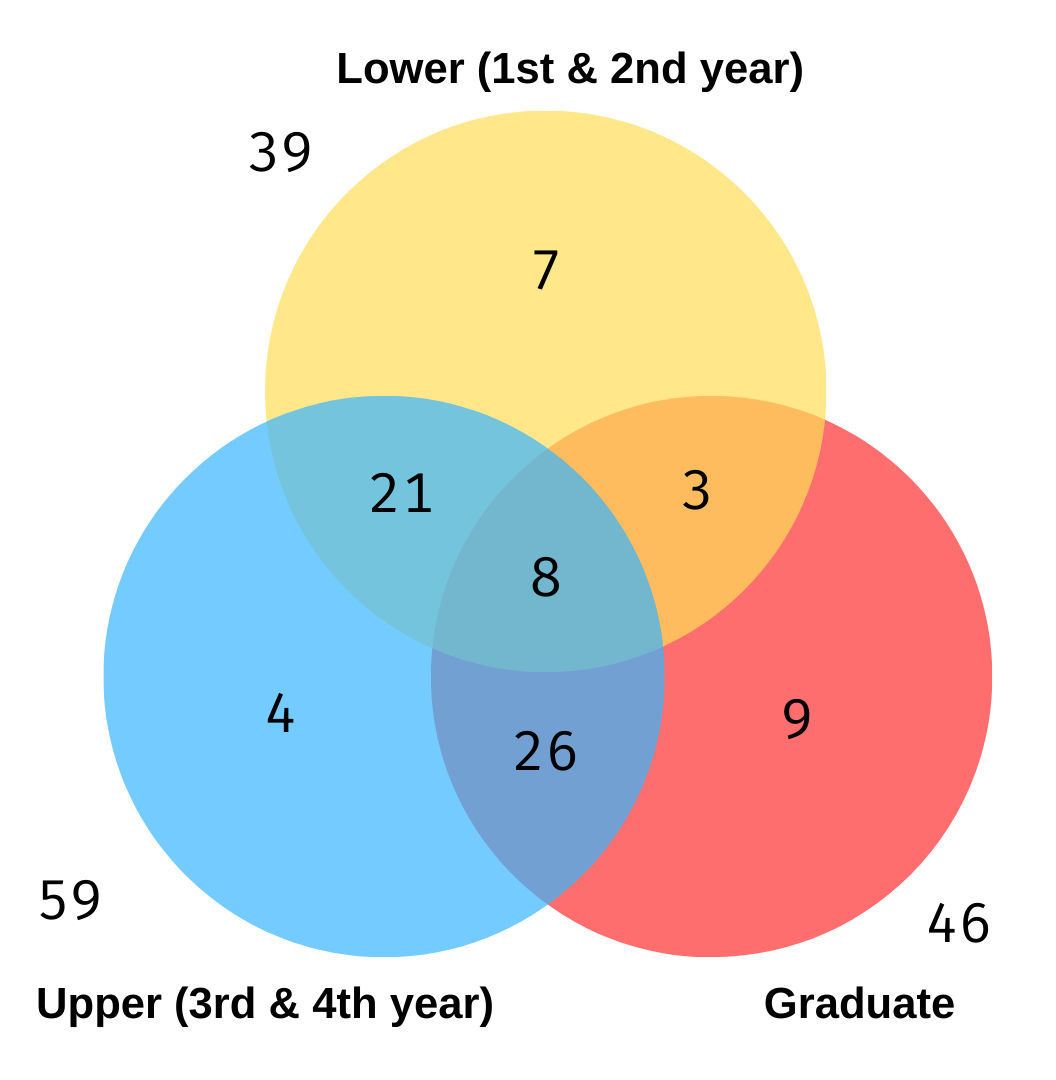}
    \caption{Distribution of Responses by Teaching Grade Level}
    \label{fig:grade}
\end{figure}

Most respondents reported teaching at multiple course levels, with upper-level undergraduate courses (3rd/4th year) being the most common at 73.8\%. More than half (57.5\%) also teach graduate-level courses, while nearly half (48.8\%) teach in lower-level undergraduate courses (1st/2nd year). Only a small fraction (two) teach secondary-level or K-12 classes. This distribution underscores the broad coverage of faculty expertise across different academic stages.  The Venn diagram in Figure \ref{fig:grade} illustrates these overlaps in teaching responsibilities (e.g., eight respondents reported teaching at all three levels)

\setlength{\textfloatsep}{0.5pt}
\section{Adoption of PjBL as Teaching Approach}
\label{section:pjbl_adoption}
This section analyzes the frequency of PjBL adoption, explores barriers to broader use, and highlights facilitators supporting its effective integration.

\subsection{PjBL Adoption Frequency}
Respondents have adopted PjBL with varying frequencies. Among the faculty, 37.5\% always implement it in all their courses, 36.3\% in most courses, and 21.3\% occasionally apply it in a few courses. A small minority (5\%) have never adopted PjBL. However, the high adoption rate may be influenced by selection bias in participant recruitment, as faculty already engaged with PjBL may have been more likely to respond. A chi-square test found no significant dependence between background factors and PjBL adoption.

Feedback on PjBL adoption frequency highlights a complex interplay of selectivity, institutional influence, uncertainty, barriers, and personal philosophy. Educators often adopt PjBL selectively, with one noting, ``I apply PjBL in a course that I believe lends itself to it because of the subject content and the material to be taught," while another emphasized its use in specific courses, saying, ``I have taught many courses, but I've only used PjBL in my current software engineering course." Institutional context shapes adoption, as reflected in the comment, ``I have been following existing course structures... These lectures included projects, so I followed suit." Personal teaching philosophy drives adoption rate for some, with one educator sharing, ``I don't always teach courses using PBL, but I think about experiential learning a lot because I also ran our senior capstone program". This feedback underscores the multifaceted nature of PjBL adoption. Uncertainty about PjBL’s definition complicates its use, with one respondent admitting, ``I haven't dug into exactly what `counts' as PBL, so I might be overestimating my adoption rate".

\subsection{PjBL Adoption Barriers}
Table \ref{tab:adoption_barriers} presents the percentage of respondents who identified with four primary categories of adoption barriers. A chi-square test found no significant dependence between background factors and barriers. Below are themes from the open-ended responses.
\begin{table}[tb]
\caption{Primary Barriers Preventing Adoption of PjBL}
\begin{center}
\begin{tabular}{|p{2.875in}|c|}
\hline
\textbf{Barrier}& \%\\
\hline
\textbf{Beliefs \& Perceptions}: Shifting from traditional teaching methods to PjBL feels inefficient or impractical for my course.  & 22.5\%\\
\hline
\textbf{Institutional Factors}: Institutional culture does not encourage / incentivize adoption, via, e.g., funding, TAs, and recognition. & 25.0\%\\
\hline
\textbf{Project Scoping and Design/Project Selection}: Designing projects that balance complexity with student skill levels is challenging. Difficult to find high-quality, adaptable PjBL projects. & 68.8\%\\
\hline
\textbf{Planning and Management of the Learning Process (Implementation, Student Support, \& Assessment)}: Managing student teamwork and collaboration in PjBL is difficult. Limited time for mentoring and a lack of technical support hindered implementation. Assessing open-ended, collaborative projects fairly is challenging. & 78.8\%\\
\hline

\end{tabular}
\label{tab:adoption_barriers}
\end{center}
\end{table}

\subsubsection{Beliefs \& Perceptions}
Some faculty expressed concerns about the overall suitability of PjBL for their courses. For instance, one respondent observed that ``if there is some content that is required knowledge, PBL feels like it can lead to potentially skipping pieces of that [knowledge]," suggesting that some educators fear essential content might be neglected. Additionally, another comment captured the sentiment that PjBL can be ``high risk, high reward", acknowledging its potential benefits while highlighting worries about students who might struggle. There was also a noted resistance to deviating from traditional methods, as one educator remarked on the ``discomfort when not using the full class time for lecture even though students learn a lot using project time during assigned class time. getting past the idea that lecturing is the only way to teach." Another faculty member shared a more optimistic stance, stating that ``There are no real barriers; from my point of view, there are only good excuses for not using this option. The current threat is AI, and the challenge is how to integrate it into the process."

\subsubsection{Institutional Factors} 
Responses pointed to a lack of institutional support as a major barrier. One educator explained that when attempting to integrate PjBL across subjects in the same semester, ``the institutions and also colleagues don't facilitate this challenge," another explained ``Just need better technology and institutional support. I do what [my institution] thinks I’m worth at this point and no more.", indicating insufficient backing. Resource limitations were also a recurring theme, with one comment stating simply ``resources." Concerns about teacher training and infrastructure were evident in remarks such as the ``ability to structure projects by teachers who do not necessarily have training and/or qualifications in the project-based learning methodology," and challenges related to large class sizes and inadequate support were underscored by the observation that ``Large class sizes and lack of TA funding are the primary barriers." Furthermore, the physical constraints of learning spaces were highlighted when an educator noted, ``I'm often assigned a large lecture hall since it is the only one that can fit 100+ students... it is awkward to have students form groups and work together." Another pointed out the lack of specific professional support:  ``... I have zero professional support on campus for feedback or revision on project ideas. The only test suite is trying a project with students and discovering the problems in real time."

\subsubsection{Project Scoping and Design/Project Selection}
Respondents emphasized the difficulty of crafting projects that are engaging and \emph{suitably} challenging. One educator remarked, ``I find PjBL easier to apply than traditional methods, but it requires a lot of thought about project design." The challenge of sourcing or creating real-world projects was also noted, as one comment mentioned that ``finding appropriate projects, particularly real-world projects, is really hard work." Additional concerns included ensuring that projects are challenging enough for every student, with one respondent emphasizing the need for ``ensuring the project contains enough challenge for each student in a team to develop on taught concepts from the module/s." Another barrier mentioned was that the ``Number and variety of topics that need to be covered in a course are [a] hindrance; it is more difficult to design a project format that covers everything." Project maintenance issues were also of concern ``maintenance/upkeep of the course with time may become difficult, especially as the underlying technology may change, necessitating the need to change the ‘Project’ in the PjBL."  Scheduling challenges are also an issue, as one educator states ``it usually takes time to design and fit appropriate projects of value into a 15 week course."

\subsubsection{Planning and Management of the Learning Process}
Revealed the most cited challenges, focusing on the complexities of organizing, executing, and assessing PjBL. One statement pinpoints that ``monitoring project progress is difficult." Not only do instructors struggle with oversight, but students themselves often grapple with coordinating deadlines, deliverables, and interpersonal dynamics ``The challenge for students is to manage management requirements for planning and long-term results against the team process." Some respondents highlighted challenges related to managing learning goals; one respondent pointed out that ``it can be especially difficult to manage the learning goals, especially the soft ones, since students tend to focus on the product rather than the process and the individual learning." One educator remarks that ``most undergraduates are not mature [enough for] PbL (require much more effort)... The ‘new’ (Z) generation has few social skills to work in teams", reflecting the added challenge of team-based projects among less experienced students. 
One educator noted that ``external assessment protocols are viewed as in conflict with PBL classroom practices," highlighting a tension between traditional evaluation methods and project-based approaches. The challenges of both team and individual assessments were clearly articulated in the observation that ``Team assessment is difficult. Individual assessment in [a] team is difficult. Require time and observations." Others noted that ``Fairly evaluating collaborative projects and managing teamwork can be difficult without the right strategies and tools." Managing large classes further complicates matters, as indicated by comments about ``applying it to the scale of large courses (dealing with 15 projects vs 70 projects)" and that ``large classes make PjBL challenging to implement." One respondent added that ``Groups are large. I don't have time for grading complex projects individually." Moreover, concerns about student readiness in certain courses were mentioned, as seen in the comment ``the focus is to ramp up the students to programming proficiency, so there is little time to dedicate to projects."

\subsection{PjBL Adoption Facilitators}
Table \ref{tab:adoption_facilitators} presents the percentage of respondents who identified with four primary categories of adoption facilitators. A chi-square test found no significant dependence between background factors and facilitators. Below are themes from the open-ended responses.

\begin{table}[tb]
\caption{Support/Strategies Encouraging PjBL Adoption}
\begin{center}
\begin{tabular}{|p{2.875in}|c|}
\hline
\textbf{Adoption Facilitators}& \textbf{\%}\\
\hline
Institutional support for project and course development and implementation (e.g., grants, funding, time release, TAs) & 68.8\%\\
\hline
Access to professional development in PjBL & 42.5\%\\
\hline
Availability of peer collaboration or mentoring opportunities & 42.5\%\\
\hline
Recognition for innovative teaching  & 37.5\%\\
\hline

\end{tabular}
\label{tab:adoption_facilitators}
\end{center}
\end{table}

\subsubsection{Institutional Support}
Several participants stressed the need for time release and ``less course load", with one noting, ``More time to think about innovative practices and implement them. We do have the money and support staff, but I’m drowned in work," and another adding, ``time. PBL requires far more out of class instructor prep." One commenter stated, ``It’s all about the money. More teachers would help," while another remarked, ``When we had administrative support in working with industry and open source project leads, it was much easier to sustain." Sustaining PjBL also relies on resources such as teaching assistants, as shown by comments like ``appropriate TAs. A major problem is that our TAs are neither trained nor qualified to work in a PjBL classroom" and ``lack of funding for teaching assistants (and paying them appropriately) is always the limiting factor." Overall, as one participant correctly summarizes, ``institutional support is the most critical."

\subsubsection{Access to Professional Development in PjBL}
Respondents highlighted the importance of professional development opportunities. To promote adoption, one participant suggests giving ``specialized training in active teaching-learning methodologies, including project-based learning." Furthermore, someone from a supportive department remarked, ``The SE department is excellent at supporting PjBL class structures... It would be nice to have (additional) prof dev..." indicating that even in an encouraging environment, continued professional development helps faculty implement and refine PjBL approaches. Another emphasizes the need for ``best practices of PjBL... or papers providing experience in adoption..."

\subsubsection{Availability of Peer Collaboration or Mentoring Opportunities}
Respondents also cited the significance of learning from and working with colleagues. One comment in particular stated, ``It would be nice to have (additional) ... collaboration to support instructors with less experience." Mentoring from those already versed in PjBL can ease the adoption of PjBL for newcomers and promote continuous improvement among more experienced instructors.

\subsubsection{Recognition for Innovative Teaching}
A little over one-third of the comments addressed the importance of rewarding faculty members who use PjBL. As one respondent explained, ``My opinion is to recognize the effort of teachers to implement PjBL in the subject..." This statement underscores how acknowledgment and incentives can encourage instructors to invest in innovative teaching methods.

\subsubsection{Other Adoption Facilitators}
 One educator affirmed ``the fact that it works as a teaching approach" is a facilitator on its own. One person wished for ``a continuity plan after the courses are finished," noting that it is ``frustrating to have to ‘shelve’ some good projects." Broader suggestions included ``Search ways to improve the teaching process" and ``Learn to measure its success and define success factors." One educator offers reassurance that ``with practice it becomes easier over time. It is important to have mechanisms to have students reflect on what worked (and didn't work)."

\section{New PjBL Project Design}
\label{section:project_design}
This section analyzes the frequency of designing new PjBL projects, explores barriers that hinder this process, and highlights facilitators that support faculty in this effort.

\subsection{New PjBL Project Design Frequency}
The results indicate that most faculty prefer to reuse projects rather than design new ones. The largest group (33.8\%) tends to reuse the same projects with minor tweaks, while 31.3\% reuse projects a few times before designing new ones. Only 25\% design a new project for every class that needs it, and the smallest group (10\%) reuses projects but with substantial revisions. This suggests a substantial portion still engages in iterative improvements rather than designing entirely new projects each time. A chi-square test of independence reveals a significant relationship between teaching experience and project design frequency (\(p=0.047\)). Specifically, less experienced faculty tend to avoid designing new projects, instead favoring project reuse with minor tweaks or substantial revisions. Experienced faculty tend to design new projects every semester or after a few uses. No significant dependence was found with PjBL adoption frequency or other background information.

\subsection{New PjBL Project Design Barriers}
Table \ref{tab:new_project_design_barriers} presents the percentage of respondents who identified with the primary new project design barriers. A chi-square test of independence reveals a significant relationship between new project design barriers and frequency. Specifically, faculty who tend to keep using the same projects with optional minor tweaks prefer using existing, proven projects mainly to avoid potential setbacks (\(p=0.031\)). No significant dependence was found with other background information or adoption frequency. Below are themes from the open-ended responses.
\begin{table}[tb]
\caption{Primary Barriers Preventing New PjBL Project Design}
\begin{center}
\begin{tabular}{|p{2.875in}|c|}
\hline
\textbf{Barriers}& \%\\
\hline
Scoping a new project requires too much time and effort. & 63.8\%\\
\hline
Designing and implementing the learning process is challenging & 46.3\%\\
\hline
Lack of Institutional support for project and course development (e.g., grants, funding, time release, teaching assistants)& 28.8\%\\
\hline
I struggle to come up with new project ideas & 27.5\%\\
\hline
Lack of peer collaboration or mentoring opportunities& 22.5\%\\
\hline
Perceived risk of new projects not being well-received by students& 21.3\%\\
\hline
Prefer using existing, proven projects to avoid potential setbacks.& 21.3\%\\
\hline

\end{tabular}
\label{tab:new_project_design_barriers}
\end{center}
\end{table}

\subsubsection{Scoping a new project requires too much time and effort}
Many comments highlight the significant workload in designing projects. One stressed that ``it’s a lot of work!" and that the ``time to engage externals to set [a] brief for projects is sometimes an issue." Another explained how ``The teaching load and substantial class size... does not allow much time for new or major revisions to course projects." Others discussed the importance of ``find[ing] alternative projects that require an acceptable time effort from students and allow them to assimilate the concepts and skills they should acquire," pointing to the challenge of aligning scope with student workload. Additional concerns include the ``difficulty adapting projects to different levels of students" and navigating ``rigid curricular requirements that limit experimentation with new projects." In data-centric domains, another instructor shared that they ``focus on research problems that involve data analysis, so one of the main barriers that constrains the choice of projects is access to high quality, relevant data sets."

\subsubsection{Designing and implementing the learning process is challenging}
Some point to the complexity of shaping PjBL structures that meet learning objectives. For instance, one person emphasized aligning outcomes and methods, stating, ``Learning outcomes need to be clearly identified to find the right project support. I generally use professional processes that I simplify to focus on the objectives," and noting that projects must remain ``up to date on tech in the industry and interesting for the students."

\subsubsection{Lack of Institutional support for project and course development}
Several comments point to institutional constraints. One laments that ``there is no incentive to rework courses. Rarely budget for that," while another notes PjBL is ``requires a very high touch course with few students and was only feasible for us at the fourth-year level," reflecting resource barriers. One person noted, ``lack of time that I personally can invest within my institution and besides my other tasks," emphasizing how institutional roles restrict innovation.

\subsubsection{I struggle to come up with new project ideas}
A few participants expressed the difficulty of generating fresh ideas. One remarked, ``Making new ones would be so hard." Another participant highlighted that estimating project complexity is difficult, especially when ``the customer is from industry, or the project topic is sourced from academia (too 'researchy'—no clear requirements)." One comment explains, ``over time, we run out of ideas for new projects."

\subsubsection{Perceived risk of new projects not being well-received by students}
One respondent directly cautions that ``'Perceived risk of new projects not being well-received by students' is sometimes underestimated, or at least, it was by me. We once tried a project that sounded awesome on paper, but it completely failed and on the scale of 1-5 (higher is better) used to evaluate courses here, we fell from high 4:s to high 2:s in just year, trigger a minor panic at the department". This highlights how fear of a project’s reception and potential fallout can deter instructors from innovating.

\subsubsection{No Barrier}
Some indicated they do not face major obstacles, stating, ``I don’t know of any barriers... I haven’t experienced either." Others rely on student-driven proposals or previous designs to reduce uncertainty, with one acknowledging that ``Making new [projects]... would be so hard." Another explains, ``For me, in computer graphics isn't difficult [to] design new projects," indicating that not everyone encounters the same challenges in creating new PjBL tasks. 

\subsection{New PjBL Project Design Facilitators}
Table \ref{tab:new_project_design_facilitators} presents the percentage of respondents who identified with the primary new project design facilitators. No significant dependence was found with background information or adoption frequency. Below are themes from the open-ended responses.

\begin{table}[tb]
\caption{New Project Design Facilitators}
\begin{center}
\begin{tabular}{|p{2.875in}|c|}
\hline
\textbf{Facilitators}& \% \\
\hline
Institutional support (e.g., grants, funding, time release, TAs)& 58.8\%\\
\hline
Leveraging collaboration with industry partners& 46.3\%\\
\hline
Professional development in PjBL& 38.8\%\\
\hline
Creating new projects as by-products of prior completed research projects  & 33.8\%\\
\hline

\end{tabular}
\label{tab:new_project_design_facilitators}
\end{center}
\end{table}

\subsubsection{Institutional support}
Some comments stressed the need for institutional support in designing new PjBL projects. One person put it as, ``time, resources, [and] interaction with other institutions," as a key facilitator. Another remarked, ``very little institutional support—limited to finding a sharp TA to develop project work."

\subsubsection{Leveraging collaboration with industry partners}
Some participants highlighted external partnerships as a vital source of project ideas. One participant shared, ``My group has several industry partners to help us source good problems. They are absolutely invaluable in keeping the project topics fresh and exciting for students." Another comment noted the value of external expertise saying ``using peoples as guest lectures who have working in the theme at hand."

\subsubsection{Professional development in PjBL}
One comment acknowledged, ``I bet I could learn a lot from what others are doing, but I don’t know who to reach out to," illustrating the need for professional development from experienced PjBL practitioners. 

\subsubsection{Creating new projects as by-products of prior completed research projects}
Participants mentioned using completed research to spark new projects. For instance, one participant noted, ``I adapt completed research projects in 3D visualization and serious games." One also states ``I don’t have difficulty to design new projects, I use recent projects done by my group, adapting the complexity to students." 

\subsubsection{Peer collaboration}
Collaborative efforts among peers were seen as helpful for sharing methods and infrastructures. One participant explained, ``peer mentoring/collaboration on PjBL infrastructure and techniques." Another mentioned, ``Have not designed PjBL projects yet but thinking of tweaking/creating project for the course by working with peers and incorporating the changes in underlying technology." One comment suggests that ``a good strategy is to try to engage other fellow teachers to carry out collaborative work". One participant emphasized the importance of open-source environments in PjBL, stating, ``the only real PjBL I have done involved creating individual or small group projects within larger open source projects. The support of the open source project maintainers was critical."

\subsubsection{Other Reflections}
One reflected on their instructional experience, explaining, ``My experience in developing and advising undergraduate and graduate theses allows me to help my students formulate their projects according to their potential." One participant articulated a broader teaching philosophy, stating, ``Time is short, and we have to simulate (simplify, abstract) real professional situations. Do not fear project failure. School is a professional sand-box. You learn from your difficulties or wrong choices." One participant suggested ``design your projects as a product line to allow variability and reuse." An alternative approach was suggested to design a new PjBL project as ``extensions of homework, smaller assignments," indicating incremental strategies. 

\section{Existing Project Adoption}
\label{section:project_adoption}
This section analyzes the frequency of adopting existing PjBL projects, explores barriers that hinder this process, and highlights facilitators that support faculty in this effort.

\subsection{Existing PjBL Project Adoption Frequency}
Faculty members exhibit different approaches to adopting existing projects for PjBL. The largest proportion (37.5\%) reported tending to keep using the same projects with only minor tweaks. Another 32.5\% indicated usually switching to a different project after using the same one a few times. Meanwhile, 18.8\% reported they adopted a different project every semester. A smaller percentage (11.3\%) reported that they kept using the same projects but with substantial revisions. A chi-square test of independence reveals a significant relationship between designing new projects and adopting existing ones (\(p=0.00\)). Specifically, faculty who design new projects are likely to adopt existing ones from other sources. No significant dependence was found with other background information or PjBL adoption frequency. 

\subsection{Existing PjBL Project Adoption Barriers}
Table \ref{tab:existing_project_adoption_barriers} presents the percentage of respondents identifying with the primary barriers to adopting existing projects. No significant dependence was found with background information or adoption frequency. Below are themes from the open-ended responses.
\begin{table} [tb]
\caption{Primary Barriers Preventing Existing PjBL Project Adoption}
\begin{center}
\begin{tabular}{|p{2.875in}|c|}
\hline
\textbf{Barriers}& \%\\
\hline
Difficulty in finding projects that align with desired learning outcomes & 50.0\%\\
\hline
Incorporating existing projects into the learning process is challenging, as I don't have sufficient time to review, adapt, and implement them. & 50.0\%\\
\hline
Lack of Institutional support (e.g., funding, time release, TAs)& 41.3\%\\
\hline
Lack of peer collaboration or mentoring opportunities& 21.3\%\\
\hline
Perceived risk of new projects not being well-received by students& 20.0\%\\
\hline
Lack of motivation to train myself to adopt existing projects & 8.8\%\\
\hline

\end{tabular}
\label{tab:existing_project_adoption_barriers}
\end{center}
\end{table}

\subsubsection{Difficulty in finding projects}
About half of the participants expressed difficulty in finding PjBL projects that align with their learning outcomes. One echoes, ``I would like to find a database with several projects so that I can use them and share the results with the community."

\subsubsection{Incorporating existing projects is challenging, as there is insufficient time to review, adapt, and implement them}
Time constraints were cited in multiple comments, including ``just too time consuming," ``the greatest challenge is finding the time," and ``I have too many other tasks on my plate to run my classes as is," indicating that participants find it difficult to integrate existing projects effectively.

\subsubsection{Lack of Institutional support} 
One participant noted ``lack of infrastructure," which suggests the need for more institutional assistance to facilitate project adoption.

\subsubsection{Lack of peer collaboration or mentoring opportunities}
Finding and maintaining peer support emerged as a challenge in the comment, ``The greatest challenge is finding the time to build new projects and to find peer support for learning new techniques," highlighting a desire for more colleague interaction in PjBL efforts.

\subsubsection{Lack of motivation}
Few participants expressed a lack of motivation to adopt existing projects. One participant referenced this issue by saying, ``lack of time, not just motivation," suggesting that while motivation is a factor, time constraints are the primary obstacle. Other remarks indicated personal preferences or context-specific reasons for not updating projects. For instance, ``It's a choice... It's not a barrier, but I keep projects for a while," ``lack of need for new unique projects," and ``I find it easy and interesting to find new projects.".

\subsubsection{Evaluation}
A participant pointed out the challenge in evaluating the effectiveness of the adopted project, saying ``difficulty evaluating the effectiveness of a new project before fully implementing it." This highlights the uncertainty instructors face when adopting a project without having immediate evidence of its impact.

\subsubsection{No barrier}
Some comments express the lack of barriers to adopting existing projects. One simply expresses ``I don't perceive barriers to adopting new projects". Another mentioned ``since our projects always involved industry mentors, the new projects came from new mentors or previous mentors with new project ideas."

\subsection{Existing PjBL Project Adoption Facilitators}
Table \ref{tab:existing_project_adoption_facilitators} presents the percentage of respondents who identified with the primary existing project adoption facilitators. No significant dependence was found with background information or adoption frequency. Below are themes from the open-ended responses.

\begin{table} [tb]
\caption{Existing Project Adoption Facilitators}
\begin{center}
\begin{tabular}{|p{2.875in}|c|}
\hline
\textbf{Facilitators}& \%\\
\hline
Institutional support (e.g., grants, funding, time release, TAs)& 51.3\%\\
\hline
Borrowing projects from fellow instructors  & 46.3\%\\
\hline
Professional development on PjBL & 40.0\%\\
\hline

\end{tabular}
\label{tab:existing_project_adoption_facilitators}
\end{center}
\end{table}

\subsubsection{Institutional Support}
Instructors stressed the need for various forms of institutional backing. One simply stated a need for ``Time to come-up with new ideas/projects," echoing the need for time release. Another reiterated the broader need for institutional backing, ``I have not been presented [with] ... institutional support". One credits a supportive environment at their institution in facilitating the sharing of projects, saying they are ``grateful...for creating an environment where PBL can flourish."

\subsubsection{Borrowing Projects from Fellow Instructors}
Several respondents underscored the value of collegial resource-sharing. One participant remarked, ``Borrowing is always good! If I could wave a wand, I'd make all the walls (Canvas and its many clones) transparent." Another emphasized, ``We certainly need to do a better job of sharing resources within and between institutions," highlighting a broader call for open exchange of materials. Similarly, another praised ``collaboration with other peers," reinforcing the importance of an environment where instructors freely borrow, adapt, and improve one another’s project ideas. One respondent also noted the importance of ``access to communities of practice or forums where experiences and proven projects are shared."

\subsubsection{Professional Development on PjBL}
One part-time professor lamented, ``as an adjunct, I have not been presented opportunities for prof dev..." This comment underscores a desire for dedicated PjBL workshops, seminars, or mentoring that could equip instructors to adopt existing PjBL projects.

\subsubsection{Other Facilitators}
One participant reflected, stating, ``strangely I just realized that I'd always designed my own projects." One suggested adapting projects ``taken from past or ongoing research.". Some comments suggest allowing students to come up with their own ideas, saying, ``The student teams create their own project (after instructor approval). This is very successful." Another also suggested ``Ask the students to decide, I check feasibility and approve the topic." Some suggest collaborating with industry, saying, ``support from industrial clients" and ``finding new willing mentors from industry."

\section{Discussion and Limitation}
\label{section:discussion}
The findings of this study highlight the complex interplay of institutional, individual, and pedagogical factors influencing faculty adoption of PjBL in computing education, aligning closely with Rogers' DOI theory \cite{rogers2014diffusion}. While many faculty recognize the pedagogical value of PjBL and report occasional to frequent use, the actual implementation is often constrained by barriers such as limited time, lack of institutional support, project scoping challenges, and difficulties in managing the learning process. These barriers particularly underscore the DOI attribute of complexity, manifesting strongly in large classes and among faculty without adequate teaching assistant support. Facilitators such as institutional recognition, professional development, and peer collaboration were found to be critical enablers, underscoring the importance of systemic and community-based support structures, which mirror DOI's emphasis on supportive social systems. On the other hand, faculty who succeed in sustaining or evolving their PjBL practices often draw upon prior research, industry partnerships, or peer collaboration to facilitate project design.

Several limitations to this study must be acknowledged. First, the high adoption rates reported may reflect a self-selection bias in our sample, as instructors already interested or invested in PjBL were more likely to respond. This could lead to overrepresenting positive experiences and underreporting challenges faced by less engaged or skeptical faculty. Second, the data is based on self-reported survey responses, which are susceptible to human error and noise, like all other surveys. Lastly, while the qualitative data enriches the findings, it likely represents only a subset of voices and may not fully capture the depth of institutional variations. Despite these limitations, the findings of this paper remain valuable as an initial set of observations on factors that impact PjBL adoption, providing a foundation for future research.

\section{Conclusion and Implication}
\label{section:conlcusion}
This study contributes to the growing body of research by illuminating the barriers and enablers impacting project-based learning adoption among computing faculty. Through a mixed-methods survey of 80 instructors, we identified key barriers, including time constraints, difficulties in project design and implementation, and insufficient institutional support that hinder widespread and sustainable PjBL implementation. At the same time, the results point to actionable strategies, such as fostering institutional incentives, providing professional development, and enabling peer collaboration, which can significantly ease adoption.

As computing education evolves to better align with industry and real-world practices, integrating and sustaining high-quality PjBL experiences will be critical for preparing students with the skills they need to thrive. The implications of our findings point to the need to scale PjBL in computing education by ensuring support goes beyond individual faculty efforts and extends to institutional infrastructure, policy, and culture. Institutions, funding agencies, and academic leaders can all play a pivotal role and should consider policies that associate PjBL adoption with reward by creating an environment that supports faculty with time, resources, and collaborative opportunities.

\section*{Acknowledgment}
This work was partially supported by the U.S. National Science Foundation Awards DUE-2111318 and DUE-2515174.

\balance

\bibliographystyle{IEEEtran}
\bibliography{IEEEabrv,main}

\begin{thebibliography}{10}
\providecommand{\url}[1]{#1}
\csname url@samestyle\endcsname
\providecommand{\newblock}{\relax}
\providecommand{\bibinfo}[2]{#2}
\providecommand{\BIBentrySTDinterwordspacing}{\spaceskip=0pt\relax}
\providecommand{\BIBentryALTinterwordstretchfactor}{4}
\providecommand{\BIBentryALTinterwordspacing}{\spaceskip=\fontdimen2\font plus
\BIBentryALTinterwordstretchfactor\fontdimen3\font minus \fontdimen4\font\relax}
\providecommand{\BIBforeignlanguage}[2]{{%
\expandafter\ifx\csname l@#1\endcsname\relax
\typeout{** WARNING: IEEEtran.bst: No hyphenation pattern has been}%
\typeout{** loaded for the language `#1'. Using the pattern for}%
\typeout{** the default language instead.}%
\else
\language=\csname l@#1\endcsname
\fi
#2}}
\providecommand{\BIBdecl}{\relax}
\BIBdecl

\bibitem{bell2010project}
S.~Bell, ``Project-based learning for the 21st century: Skills for the future,'' \emph{The clearing house}, vol.~83, no.~2, pp. 39--43, 2010.

\bibitem{kokotsaki2016project}
D.~Kokotsaki, V.~Menzies, and A.~Wiggins, ``Project-based learning: A review of the literature,'' \emph{Improving schools}, vol.~19, no.~3, pp. 267--277, 2016.

\bibitem{gupta2022impact}
C.~Gupta, ``The impact and measurement of today’s learning technologies in teaching software engineering course using design-based learning and project-based learning,'' \emph{IEEE Transactions on Education}, vol.~65, no.~4, pp. 703--712, 2022.

\bibitem{fu2018teaching}
Y.~Fu, L.~P. Reina, and P.~Brockmann, ``Teaching global software engineering: Experience report comparing distributed, virtual collaborative courses at the bachelor's and master's degree levels,'' in \emph{Proceedings of the 3rd European Conference of Software Engineering Education}, 2018, pp. 34--38.

\bibitem{fioravanti2018integrating}
M.~L. Fioravanti, B.~Sena, L.~N. Paschoal, L.~R. Silva, A.~P. Allian, E.~Y. Nakagawa, S.~R. Souza, S.~Isotani, and E.~F. Barbosa, ``Integrating project based learning and project management for software engineering teaching: An experience report,'' in \emph{Proceedings of the 49th ACM technical symposium on computer science education}, 2018, pp. 806--811.

\bibitem{tubino2021reforming}
L.~Tubino, J.-G. Schneider, A.~Cain, D.~Thiruvady, and C.~Ranaweera, ``Reforming assessment: challenges beyond design,'' in \emph{2021 IEEE/ACM 43rd International Conference on Software Engineering: Software Engineering Education and Training (ICSE-SEET)}.\hskip 1em plus 0.5em minus 0.4em\relax IEEE, 2021, pp. 78--88.

\bibitem{papadopoulos2012students}
P.~M. Papadopoulos, I.~G. Stamelos, and A.~Meiszner, ``Students’perspectives on learning software engineering with open source projects-lessons learnt after three years of program operation,'' in \emph{International Conference on Computer Supported Education}, vol.~2.\hskip 1em plus 0.5em minus 0.4em\relax SCITEPRESS, 2012, pp. 313--322.

\bibitem{suleiman2024providing}
A.~D. Suleiman, D.~Shepherd, J.~DeWaters, Y.~Liu, and D.~Hou, ``Providing technical support to sustain student motivation and engagement in software engineering project-based learning,'' in \emph{2024 IEEE Frontiers in Education Conference (FIE)}.\hskip 1em plus 0.5em minus 0.4em\relax IEEE, 2024, pp. 1--9.

\bibitem{daun2016project}
M.~Daun, A.~Salmon, T.~Weyer, K.~Pohl, and B.~Tenbergen, ``Project-based learning with examples from industry in university courses: an experience report from an undergraduate requirements engineering course,'' in \emph{2016 IEEE 29th International Conference on Software Engineering Education and Training (CSEET)}.\hskip 1em plus 0.5em minus 0.4em\relax IEEE, 2016, pp. 184--193.

\bibitem{fernandes2020achieving}
J.~P. Fernandes, R.~Ara{\'u}jo, and M.~Zenha-Rela, ``Achieving scalability in project based learning through a low-code platform,'' in \emph{Proceedings of the XXXIV Brazilian Symposium on Software Engineering}, 2020, pp. 710--719.

\bibitem{mason2019collaboration}
T.~Mason, A.~Gavrilovska, and D.~A. Joyner, ``Collaboration versus cheating: Reducing code plagiarism in an online ms computer science program,'' in \emph{Proceedings of the 50th ACM Technical Symposium on Computer Science Education}, 2019, pp. 1004--1010.

\bibitem{marti2006pbl}
E.~Mart{\'\i}, D.~Gil, and C.~Juli{\`a}, ``A pbl experience in the teaching of computer graphics,'' in \emph{Computer Graphics Forum}, vol.~25, no.~1.\hskip 1em plus 0.5em minus 0.4em\relax Wiley Online Library, 2006, pp. 95--103.

\bibitem{rogers2014diffusion}
E.~M. Rogers, A.~Singhal, and M.~M. Quinlan, ``Diffusion of innovations,'' in \emph{An integrated approach to communication theory and research}.\hskip 1em plus 0.5em minus 0.4em\relax Routledge, 2014, pp. 432--448.

\bibitem{meng2023tackle}
N.~Meng, Y.~Dong, D.~Roehrs, and L.~Luan, ``Tackle implementation challenges in project-based learning: a survey study of pbl e-learning platforms,'' \emph{Educational technology research and development}, vol.~71, no.~3, pp. 1179--1207, 2023.

\bibitem{rehman_trends}
S.~U. Rehman, ``Trends and challenges of project-based learning in computer science and engineering education,'' in \emph{Proceedings of the 15th International Conference on Education Technology and Computers}, ser. ICETC '23.\hskip 1em plus 0.5em minus 0.4em\relax Association for Computing Machinery, 2024, p. 397–403.

\bibitem{sahin2006detailed}
I.~Sahin, ``Detailed review of rogers' diffusion of innovations theory and educational technology-related studies based on rogers' theory.'' \emph{Turkish Online Journal of Educational Technology-TOJET}, vol.~5, no.~2, pp. 14--23, 2006.

\bibitem{rees2019opportunities}
D.~G. Rees~Lewis, E.~M. Gerber, S.~E. Carlson, and M.~W. Easterday, ``Opportunities for educational innovations in authentic project-based learning: understanding instructor perceived challenges to design for adoption,'' \emph{Educational technology research and development}, vol.~67, pp. 953--982, 2019.

\bibitem{ahmadSLR}
A.~D. Suleiman, D.~Hou, Y.~Liu, J.~DeWaters, D.~Shepherd, and J.~G. de~Souza, ``Systematic literature review on project-based learning in computing education,'' 2025, accepted for publication, expected publication date: 2025.

\bibitem{mchugh2013chi}
M.~L. McHugh, ``The chi-square test of independence,'' \emph{Biochemia medica}, vol.~23, no.~2, pp. 143--149, 2013.

\bibitem{Braun01012006}
V.~Braun and V.~Clarke, ``Using thematic analysis in psychology,'' \emph{Qualitative Research in Psychology}, vol.~3, no.~2, pp. 77--101, 2006.

\end{thebibliography}

\end{document}